\documentclass[11pt]{article}\textwidth 6.5in\textheight 9in
\usepackage{amssymb}\usepackage[colorlinks]{hyperref}\usepackage{color}
	\usepackage{stmaryrd}\usepackage{mathrsfs}
		\usepackage{graphicx}
	\usepackage{amsmath}
\usepackage{caption}
    \usepackage{subcaption}
    \usepackage{listings}
    \usepackage{verbatim}
    \usepackage{braket}
 
\begin{document}
\setlength{\captionmargin}{27pt}
\newcommand\hreff[1]{\href {http://#1} {\small http://#1}}
\newcommand\trm[1]{{\bf\em #1}} \newcommand\emm[1]{{\ensuremath{#1}}}
\newcommand\prf{\paragraph{Proof.}}\newcommand\qed{\hfill\emm\blacksquare}

\newtheorem{thr}{Theorem} 
\newtheorem{lmm}{Lemma}
\newtheorem{cor}{Corollary}
\newtheorem{con}{Conjecture} 
\newtheorem{prp}{Proposition}

\newcommand\QC{\mathbf{QC}} 
\newcommand\C{\mathbf{C}} 
\renewcommand\H{\mathbf{H}}

\newtheorem{blk}{Block}
\newtheorem{dff}{Definition}
\newtheorem{asm}{Assumption}
\newtheorem{rmk}{Remark}
\newtheorem{clm}{Claim}
\newtheorem{exm}{Example}

\newcommand\Ks{\mathbf{Ks}} 
\newcommand{\ab}{a\!b}
\newcommand{\yx}{y\!x}
\newcommand{\yux}{y\!\underline{x}}

\newcommand\floor[1]{{\lfloor#1\rfloor}}\newcommand\ceil[1]{{\lceil#1\rceil}}

\newcommand{\bmu}{\boldsymbol{\mu}}

\newcommand{\lea}{<^+}
\newcommand{\gea}{>^+}
\newcommand{\eqa}{=^+}

\newcommand{\lel}{<^{\log}}
\newcommand{\gel}{>^{\log}}
\newcommand{\eql}{=^{\log}}

\newcommand{\F}{\mathbf{F}}
\newcommand{\E}{\mathbf{E}}
\newcommand{\lem}{\stackrel{\ast}{<}}
\newcommand{\gem}{\stackrel{\ast}{>}}
\newcommand{\eqm}{\stackrel{\ast}{=}}

\newcommand\edf{{\,\stackrel{\mbox{\tiny def}}=\,}}
\newcommand\edl{{\,\stackrel{\mbox{\tiny def}}\leq\,}}
\newcommand\then{\Rightarrow}

\newcommand\ml{\underline{\mathbf m}}

\renewcommand\chi{\mathcal{H}}
\newcommand\km{{\mathbf {km}}}\renewcommand\t{{\mathbf {t}}}
\newcommand\KM{{\mathbf {KM}}}\newcommand\m{{\mathbf {m}}}
\newcommand\md{{\mathbf {m}_{\mathbf{d}}}}\newcommand\mT{{\mathbf {m}_{\mathbf{T}}}}
\newcommand\K{{\mathbf K}} \newcommand\I{{\mathbf I}}

\newcommand\II{\hat{\mathbf I}}
\newcommand\Kd{{\mathbf{Kd}}} \newcommand\KT{{\mathbf{KT}}} 
\renewcommand\d{{\mathbf d}} 
\newcommand\D{{\mathbf D}}
\newcommand\Tr{\mathrm{Tr}}
\newcommand\w{{\mathbf w}}
\newcommand\Cs{\mathbf{Cs}} \newcommand\q{{\mathbf q}}
\newcommand\St{{\mathbf S}}
\newcommand\M{{\mathbf M}}\newcommand\Q{{\mathbf Q}}
\newcommand\ch{{\mathcal H}} \renewcommand\l{\tau}
\newcommand\tb{{\mathbf t}} \renewcommand\L{{\mathbf L}}
\newcommand\bb{{\mathbf {bb}}}\newcommand\Km{{\mathbf {Km}}}
\renewcommand\q{{\mathbf q}}\newcommand\J{{\mathbf J}}
\newcommand\z{\mathbf{z}}

\newcommand\B{\mathbf{bb}}\newcommand\f{\mathbf{f}}
\newcommand\hd{\mathbf{0'}} \newcommand\T{{\mathbf T}}
\newcommand\R{\mathbb{R}}\renewcommand\Q{\mathbb{Q}}
\newcommand\N{\mathbb{N}}\newcommand\BT{\{0,1\}}
\newcommand\FS{\BT^*}\newcommand\IS{\BT^\infty}
\newcommand\FIS{\BT^{*\infty}}
\renewcommand\S{\mathcal{C}}\newcommand\ST{\mathcal{S}}
\newcommand\UM{\nu_0}\newcommand\EN{\mathcal{W}}

\newcommand{\supp}{\mathrm{Supp}}

\newcommand\lenum{\lbrack\!\lbrack}
\newcommand\renum{\rbrack\!\rbrack}

\newcommand\h{\mathbf{h}}
\renewcommand\qed{\hfill\emm\square}
\renewcommand\i{\mathbf{i}}
\newcommand\p{\mathbf{p}}
\renewcommand\q{\mathbf{q}}
\title{ A Quantum Outlier Theorem}

\author {Samuel Epstein\\samepst@jptheorygroup.org}

\maketitle

\begin{abstract}
In recent results, it has been proven that all sampling methods produce outliers. In this paper, we extend these results to quantum information theory. Projectors of large rank must contain pure quantum states in their images that are outlying states. Otherwise, the projectors are exotic, in that they have high mutual information with the halting sequence. Thus quantum coding schemes that use projections, such as Schumacher compression, must communicate using outlier quantum states.
\end{abstract}

\section{Introduction}

In algorithmic information theory, the notion of an outlier is modelled using the randomness deficiency. The model is defined by a probability $p$, over natural numbers, and the data point $x$ is a natural number. The randomness deficiency is formally defined as

$$\d(x|p) =\log \mathbf{t}_p(x),$$

where $\mathbf{t}_p$ is a universal lower computable $p$-test. It is a score of how atypical a datapoint is with respect to a model. In \cite{Gacs01}, the quantum notion of randomness deficiency was introduced. This quantum randomness deficiency measures the algorithmic atypicality of a pure or mixed quantum state $\rho$ with respect to a second quantum mixed state $\sigma$. It is defined by 

$$\d(\rho|\sigma) = \log \Tr \rho \textbf{t}_\sigma,$$

where $\textbf{t}_\sigma$ is a universal lower computable Hermitian matrix such that $\Tr\sigma \textbf{t}_\sigma\leq1$. The density matrix $\sigma$ is assumed to be computable. Mixed states are used to model random mixtures $\{p_i\}$ of pure states $\{\ket{\psi_i}\}$, so quantum randomness deficiency is a score of how atypical a quantum state is with respect to a mixture. See Section \ref{sec:conv} for more motivation for the definition of quantum randomness deficiency. In \cite{EpsteinClone19}, quantum randomness deficiency was extended to uncomputable quantum states.

What are the interesting properties of quantum randomness deficiency? In \cite{EpsteinClone19}, conservation of quantum randomness deficiency was proven over partial trace and unitary operations. With some work conservation over quantum operations can be proven. One recent result in the classical randomness deficiency case is that sampling methods produce outliers \cite{Epstein21}. There are several proofs to this result, with one of them derived from the fact that large sets of natural numbers with low randomness deficiencies are exotic, in that they have high mutual information with the halting sequence.

In this paper, we prove a quantum version of this result. Projections of large rank must contain pure quantum states in their images that are outlying states. Otherwise, the projections are exotic, in that they have high mutual information with the halting sequence. Thus quantum coding schemes that use projections, such as Schumacher compression, must communicate using outlier quantum states. The classical and quantum theorems are analogous, but their proofs are very different!

\section{Conventions}
\label{sec:conv}
For positive real function $f$, $\lea f$, $\gea f$, and $\eqa f$ is used to represent $< f+O(1)$, $>f+O(1)$, and $=f\pm O(1)$. For the nonnegative real function $f$, the terms ${\lel}f$, ${\gel} f$, and ${\eql}f$ represent the terms ${<}f{+}O(\log(f{+}1))$, ${>}f{-}O(\log(f{+}1))$, and ${=}f{\pm}O(\log(f{+}1))$, respectively. 

Let $\K(x)$ be the prefix free Kolmogorov complexity. We use $\I(x;\ch)=\K(x)-\K(x|\ch)$ to be the amount of information that the halting sequence $\ch\in\IS$ has about $x\in\FS$. A probability is elementary if it has finite support and all its values are rational. The deficiency of randomness of a string $x$ with respect to an elementary probability mesaure $Q$ is $\d(x|Q) = \floor{-\log Q(x)}-\K(x|\langle Q\rangle)$. The stochasticity of a string is $\Ks(x) = \min_Q\{\K(Q)+3\log\max\{\d(x|Q),1\}\}$.
\begin{lmm}[\cite{Epstein21,Levin16}]
\label{lmm:ks}
$\Ks(x)\lel \I(x;\ch)$.
\end{lmm}

We use $\mathcal{H}_n$ to denote a Hilbert space with $n$ dimensions, spanned by bases $\ket{\beta_1},\dots,\ket{\beta_n}$. A qubit is a unit vector in the Hilbert space $\mathcal{H}_2$, spanned by vectors $\ket{0}$, $\ket{1}$. To model $n$ qubits, we use a unit vector in $\mathcal{H}_{2^n}$, spanned by basis vectors $\ket{x}$, where $x$ is a string of size $n$. 

A pure quantum state $\ket{\psi}$ of length $n$ is a unit vector in $\mathcal{H}_{2^n}$. Its corresponding element in the dual space is denoted by $\bra{\phi}$. The conjugate transpose of a a matrix $A$ is $A^*$. $\Tr$ is used to denote the trace of a matrix. 
Projection matrices are Hermitian matrices with eigenvalues in $\{0,1\}$. 

A complex matrix $A$ is elementary if its entries are complex numbers with rational coefficients and can be encoded as $\langle A\rangle$, and has a Kolmogorov complexity $\K(A)$.

For Hermitian matrices, $\sigma \leq\rho$ iff $\rho-\sigma$ is Hermitian. We say program $q\in\FS$ lower computes Hermitian matrix $\sigma$ if, given as input to universal Turing machine $U$, the machine $U$ reads  $\leq\|q\|$ bits and outputs, with or without halting, a sequence of elementary semi-density matrices $\{\sigma_i\}$ such that $\sigma_{i}\leq \sigma_{i+1}$ and $\lim_{i\rightarrow\infty}\sigma_i = \sigma$. A matrix $T$ is lower computable if there is a program that lower computes it. Its complexity is $\K(T) = \min \{\K(q) :q\textrm{ lower computes }T\}$. Given a density matrix $\sigma$, a $\sigma$-test is a lower computable Hermitian matrx $T$ such that $\Tr T\sigma=1$. If $\sigma$ is computable, there exists a universal $\sigma$ test $\mathbf{t}_\sigma$, that is lower computable relative to the number of qubits $n$, $\Tr\sigma\textbf{t}_\sigma\leq 1$, and for every lower computable $\sigma$ test $T$, $O(1)\t_{\sigma} > 2^{-\K(T|\sigma)}T$. This universal test can be computed in the standard way, analagously to the classical case (see \cite{Gacs21}).
\begin{dff}[Quantum Randomness Deficiency]
For mixed states $\sigma$ and $\rho$, $\d(\rho|\sigma)=\log \Tr\,\mathbf{t}_{\sigma}\rho$.
\end{dff}
The quantum randomness deficiency, among other interpretations, is score of how typical a pure state is with respect to an algorithmically generated quantum source. Indeed, suppose there is a computable probability $P$ over encodings of elementary orthogonal pure states $\{\langle \ket{\psi_i}\rangle\}$ of orthogonal pure states $\{\ket{\psi_i}\}$, with corresponding density matrix $\sigma = \sum_iP(\langle \ket{\psi_i}\rangle)\ket{\psi_i}\bra{\psi_i}$. Then there is a lower-computable $\sigma$-test $T=\sum_i2^{\d(\langle \ket{\psi_i}\rangle|P)}\ket{\psi_i}\bra{\psi_i}$ with $O(1)\t_\sigma>T$. Thus $\d(\ket{\psi_i}|\sigma)\gea \d(\langle \ket{\psi_i}\rangle|P)$, giving high scores to pure states $\ket{\psi_i}$ which are atypical of the source. In general the $\d(\ket{\phi}|\sigma)$ score for arbitrary $\ket{\phi}$ will be greater than a combination of $\d(\cdot|P)$ scores, with
$\d(\ket{\phi}|\sigma)\gea \log \sum2^{\d(\langle \ket{\psi_i}\rangle|P)}|\braket{\phi|\psi_i}|^2$.\newpage

\section{Results}
\begin{thr}
\label{thr}
Relativized to an $n$ qubit mixed state $\sigma$, for elementary $2^m$ rank  projector $P$,\\ $3m-2n\lel\max_{\ket{\phi}\in\mathrm{Image}(P)}\d(\ket{\phi}|\sigma)+\I(\langle P\rangle ;\ch)$.
\end{thr}
\begin{prf}
We relativize the universal Turing machine to $\langle\sigma\rangle$ and $(3m-2n)$. Thus it is effectively relativized to $m$, $n$, and $\sigma$. Let elementary probability measure $Q$ and $d\in\N$ realize $\Ks(P)$, where $d=\max\{\d(P|Q),1\}$. Without loss of generality we can assume that the support of $Q$ is elementary projections of rank $2^m$. There are $d2^{n-m+2}$ rounds. For each round we select an $\sigma$-test $T$, that is of dimension 1, $\Tr\sigma T\leq 1$, and for a certain $Q$-probability of projections $B$, $\Tr TB$ is large. We now describe the selection process.

Select a random test $T$ to be $2^{m-2}\ket{\psi}\bra{\psi}$, where $\ket{\psi}$ is an $n$ qubit state chosen uniformly from the unit sphere, with distribution $\Lambda$.
$$
\E[\Tr T\sigma] = 2^{m-2}\int \Tr\bra{\psi}\sigma\ket{\psi}d\Lambda = 2^{m-2}\Tr\sigma\int \ket{\psi}\bra{\psi}d\Lambda=2^{m-n-2}\Tr\sigma=2^{m-n-2}.
$$
Thus the probability that $T$ is a $\sigma$-test is $\geq1- 2^{m-n-2}$. Let $I_m$ be an $n$-qubit identity matrix with only the first $2^m$ diagonal elements being non-zero. Let $K_m = I-I_m$. Let $p=2^{m-n}$ and $\hat{T} = T/2^{m-2}$. For any projection $B$ of rank $2^m$,
\begin{align*}
&\Pr(\Tr B\hat{T}\leq.5p)\\
=&\Pr(\Tr I_m\hat{T}\leq.5p)\\
=&\Pr(\Tr K_m\hat{T}\geq 1-.5p)\\
\E[\Tr K_m\hat{T}] &= 1-p\\
\Pr(\Tr K_m\hat{T}\geq 1-.5p) &\leq (1-p)/(1-.5p)\\
\Pr(\Tr B\hat{T}\geq .5p)&=1-\Pr(\Tr K_m\hat{T}\geq 1-.5p)\\
 &\geq 1 - (1-p)/(1-.5p)\\
&= .5p/(1-.5p) \geq .5p\\
\Pr(\Tr BT\geq 2^{2m-n-3}) &  \geq .5p.
\end{align*}
Let $\Omega$ be the space of all matrices of the form $2^{m-2}\ket{\phi}\bra{\phi}$. Let $R$ be the uniform distribution over $\Omega$. Let $[A,B]$ be 1 if $\Tr AB>2^{2m-n-3}$, and 0 otherwise. By the above equations, for all $A\in\mathrm{Support}(Q)$, $\int_\Omega[A,B]dR(B)\geq.5p$. So $\sum_A\int_\Omega[A,B]Q(A)dR(B)\geq.5p$. For Hermitian matrix $A$, $\{A\}$ is 1 if $\Tr A\sigma\leq 1$, and 0 otherwise. So $\int_\Omega\{A\}dR(A)\geq (1-p2^{-2})$. Let $f=\max_{T}\{T\}\sum Q(A)[T,A]$. 

So
\begin{align*}
.5p &\leq \sum_A\int_\Omega[A,B]Q(A)dR(B)\\
 &=  \sum_A\int_{\Omega}\{B\}Q[A,B](A)dR(B)+
\sum_A\int_\Omega(1-\{B\})[A,B]Q(A)dR(B)
\\
  &\leq  \sum_A\int_\Omega\{B\}[A,B]Q(A)dR(B)+\int_\Omega(1-\{B\})dR(B)\\
    &\leq  \sum_A\int_\Omega\{B\}[A,B]Q(A)dR(B)+p2^{-2}\\
p/4&\leq  \sum_A\int_\Omega\{B\}[A,B]Q(A)dR(B)=\int_\Omega\left(\{B\}\sum_A[A,B]Q(A)\right)dR(B)\leq \int_{\Omega} fdR(B)\\
p/4& \leq f.
\end{align*}

Thus for each round $i$, the lower bounds on $f$ proves there exists a one dimensional matrix $T_i=2^{m-2}\ket{\psi}\bra{\psi}$ such that $\Tr T_i\sigma\leq 1$ and $\sum_R \{Q(R) : \Tr T_iR\geq 2^{2m-n-3}\}\geq p/4 = 2^{m-n-2}$. Such a $T_i$ is selected, and the the $Q$ probability is conditioned on those projections $B$ for which $[T_i,B]=0$, and the next round starts. Assuming that there are $d2^{n-m+2}$ rounds, the $Q$ measure of projections $B$ such there does not exist a $T_i$ with $[T_i,B]=1$ is 
$$\leq (1-p/4)^{d2^{n-m+2}}\leq e^{-d}.$$
Thus there exists a $T_i$ such that $[T_i,P]=1$, otherwise one can create a $Q$ test $t$ that assigns $e^d$ to all projections $B$ where there does not exist $T_i$ with $[T_i,B]=1$, and 0 otherwise. Then $t(P)=e^d$ so
$$ 1.44 d < \log t(P) \lea \d(P|Q) \lea d.$$
This is a contradiction, because without loss of generality, one can assume $d$ is large. Let $T_i=2^{m-2}\ket{\psi}\bra{\psi}$ with $[T_i,P]=1$.
Let $\ket{\phi} = P\ket{\psi}/\sqrt{\bra{\psi}P\ket{\psi}}$. So $\bra{\phi}T_i\ket{\phi}\geq 2^{2m-n-3}$ and $\ket{\phi}$ is in the image of $P$. Thus by Lemma \ref{lmm:ks},
\begin{align*}
2m-n &\lea \log \bra{\phi}T_i\ket{\phi}\\
2m-n &\lea \log \max_{\ket{\phi}\in\textrm{Image}(P)}\bra{\phi}T_i\ket{\phi}\\
2m-n &\lea \max_{\ket{\phi}\in\textrm{Image}(P)}\d(P|\sigma)+\K(T_i)\\
2m-n &\lea \max_{\ket{\phi}\in\textrm{Image}(P)}\d(P|\sigma)+(n-m)+\log d+\K(d)+\K(Q)\\
2m-n &\lea \max_{\ket{\phi}\in\textrm{Image}(P)}\d(P|\sigma)+(n-m)+\Ks(P)\\
3m-2n &\lel\max_{\ket{\phi}\in\textrm{Image}(P)} \d(P|\sigma) + \I(P;\ch).
\end{align*}
Note that due to the fact that the left hand side of the equation is $(3m-2n)$ and it has log precision, this enables one to condition the universal Turing machine  to $(3m-2n)$.
\qed
\end{prf}$  $\newpage
\subsection{Computable Projections}

Theorem \ref{thr} is in terms of elementary described projecctions and can be generalized to arbitrarily computable projections. For a matrix $M$, let $\|M\|=\max_{i,j}|M_{i,j}|$ be the max norm. A program $p\in\FS$ computes a projection $P$ of rank $\ell$ if it outputs a series of rank $\ell$ projections $\{P_i\}_{i=1}^\infty$ such that $\|P-P_i\|\leq 2^{-i}$. For computable projection operator $P$, $\I(P;\ch)=\min\{\K(p)-\K(p|\ch):p\textrm{ is a program that computes }P\}$. 
\begin{lmm}[\cite{EpsteinDerandom22}]
\label{lmm}
For partial computable $f$, $\I(f(a);\ch)\lea \I(a;\ch)+\K(f)$.
\end{lmm}
\begin{cor}
\label{cor}
Relativized to an $n$ qubit mixed state $\sigma$, for computable $2^m$ rank  projector $P$, $3m-2n\lel\max_{\ket{\phi}\in\mathrm{Image}(P)}\d(\ket{\phi}|\sigma)+\I(\langle P\rangle ;\ch)$.
\end{cor}
\begin{prf}
Let $p$ be a program that computes $P$. There is a simply defined algorithm $A$, that when given $p$ and $\sigma$, outputs $P_n$ such that $\max_{\ket{\psi}\in\mathrm{Image}(P)}\d(\ket{\psi}|\sigma)\eqa\max_{\ket{\psi}\in\mathrm{Image}(P_n)}\d(\ket{\psi}|\sigma)$. Thus by Lemma \ref{lmm}, one gets that $\I(P_n;\ch)\lea \I(P;\ch)$. The corollary follows from Theorem \ref{thr}.
\qed
\end{prf}
\section{Discussion}
In previous work \cite{Epstein21,EpsteinDerandom22,EpsteinDynamics22,EpsteinThermo23}, outliers are shown to be emergent in dynamics and the outputs of probabilistic algorithms. This was achieved by modeling outliers using the randomness deficiency function. This function has been defined over strings, infinite sequences, or points in computable metric spaces.

This work shows that the ubiquity of outliers is present in the physical realm, namely quantum mechanics. Quantum communication schemes of quantum sources that involves quantum projectors of large rank must communicate with atypical quantum states with respect to the quantum sources. Future work can look into the provable presence of algorithmic anomalies in other areas of physics, such as quantum field theory or black holes.

\end{document}